\documentclass[twocolumn,showpacs,preprintnumbers,amsmath,amssymb]{revtex4-1}

\usepackage{graphicx}
\usepackage{dcolumn}
\usepackage{bm}
\usepackage{epstopdf}
\usepackage{amssymb}
\usepackage{verbatim}
\usepackage{epsfig}
\usepackage{tikz}        
\usepackage{etoolbox}

\usepackage{lineno}
\usepackage{mwe}
\usepackage{subfigure}

\begin{document}

\newrobustcmd*{\mydiamond}[1]{\tikz{\filldraw[black,fill=#1] 
(0,0) -- (0.2cm,0.2cm) -- (0.4cm,0) -- (0.2cm,-0.2cm) -- (0,0);}}

\newrobustcmd*{\mytriangleright}[1]{\tikz{\filldraw[black,fill=#1] (0,0.2cm) -- 
(0.3cm,0) -- (0,-0.2cm);}}


\title{Cyclical lock-down and the economic activity along the pandemic of 
COVID-19} 
%

 
    \author{F.E.~Cornes}
  \affiliation{Departamento de F\'\i sica, Facultad de Ciencias Exactas y 
Naturales, Universidad de Buenos Aires,
 Pabell\'on I, Ciudad Universitaria, 1428 Buenos Aires, Argentina.}

  \author{G.A.~Frank}
  \affiliation{Unidad de Investigaci\'on y Desarrollo de las 
Ingenier\'\i as, Universidad Tecnol\'ogica Nacional, Facultad Regional Buenos 
Aires, Av. Medrano 951, 1179 Buenos Aires, Argentina.}

  \author{C.O.~Dorso}%
  \affiliation{Departamento de F\'\i sica, Facultad de Ciencias Exactas y 
Naturales, Universidad de Buenos Aires,
 Pabell\'on I, Ciudad Universitaria, 1428 Buenos Aires, Argentina.}
  \affiliation{Instituto de F\'\i sica de Buenos Aires, Pabell\'on I, 
Ciudad Universitaria, 1428 Buenos Aires, Argentina.}

 \email{codorso@df.uba.ar}

\date{\today}

\begin{abstract}
The investigation focuses on an on-off protocol for controlling the
COVID-19 widespread. The protocol establishes a working period of 4
days for all the citizens, followed by 8 days of lock-down. We further
propose splitting people into smaller groups that undergo the
on-off protocol, but shifted in time. This procedure is expected to
regularize the overall economic activity. Our results show that either the 
on-of protocol and the splitting into groups reduces the amount of infected 
people. However, the latter seems to be better for economic reasons. Our 
simulations further show that the start-up time is a key issue for the  
success of the implementation.
\end{abstract}


\maketitle

\section{\label{introduction}Introduction}

In this investigation we explore a cyclic scheme of isolation (quarantine) 
and economic activity in order to cope with a pandemic like the Covid-19 
one. The proposed scheme consists in a sequence of isolation-activity cycles 
in which a portion of the population (one third) is fully free to join its 
usual activities while the other two thirds are subject of the isolation. 
After four days the third that has been working goes into quarantine (8 
days isolation), one of the remaining thirds remains in quarantine and the 
third one change to a state of full activity. As will be shown this scheme 
(inspired in Ref.~\cite{Alon}) allows to properly control the pandemic 
evolution while keeping the economic cycle active. \\ 

The evolution of the disease is performed using a SEIR (Susceptible, 
Exposed, Infected, Recovered) compartmental model .

In Section~\ref{sec:seir} we briefly describe the SEIR model. In 
Section~\ref{sec:the_model_cyclic} we show the way in 
which the above metioned cyclic alternate scheme is implemented in the 4-8 
case.\\

\section{\label{sec:modelo}The Model}

\subsection{\label{sec:seir} The SEIR model of a single group}

In order to describe the time evolution of a given population when a (small) 
fraction of it is infected, we resort to the SEIR compartmental model. These 
kind of models consider that the individuals among the the population can be  
in one of the following possible states:

\begin{itemize}
 \item[S:] Susceptible individuals, which are not immune to the considered 
infection, and consequently, can get infected by contact with an infected 
individual.
\item[E:] Exposed, is an individual who having been in contact with I, is 
already a patient but is unable to infect other Susceptible individuals.
\item[I:] Is the infected state, the individual is able to infect other 
susceptible individuals.
\item[R:] Removed state, the individuals in this state do not participate in 
the process of epidemic evolution any more. It both comprises the individuals 
who became immune to the illness but also those who die.
\end{itemize}

In this way this quantities satisfy the following relation

\begin{equation}
 S(t)+E(t)+I(t)+R(t)=N
\end{equation}

The SEIR model appears as a more accurate model for the coronavirus COVID-19 
pandemic with respect to the SIS, SIR, etc.. This is because those individuals 
who catched the infection undergo an ``incubation'' period, through which they 
are not able to infect others. \\

The equations that describe the evolution of the infection read as follows.

\begin{equation}
\left\{\begin{array}{lcl}
\dot{s}(t) & = & -\beta i(t)s(t)\\
\dot{e}(t) & = & \beta i(t)s(t)-\sigma e(t)\\
\dot{i}(t) & = & \sigma e(t)-\gamma i(t)\\
\dot{r}(t) & = & \gamma i(t)
\end{array}\right.\label{eq:seir}
\end{equation}
    
\noindent with $s(t)=S(t)/N$, $e(t)=E(t)/N$, etc.  These are the magnitudes 
per unit individual. For the purpose of simplicity, we will consider $\beta$, 
$\sigma$ y $\gamma$ as fixed parameters. The parameter $\beta$ (infection 
rate) represents the effective mean field rate of infection, actuating on the 
product of the relative susceptible population and the relative infected 
population. It depends on intrinsic ingredients like the infectivity of the 
virus under consideration and extrinsic ones like the contact frequency.  
Besides, the parameters $\sigma$ and $\gamma$ depend exclusively on the 
illness under consideration.

One of the relevant magnitudes is the so called basic reproduction number 
$R_0$ \cite{Dushoff}
 
\begin{equation}
    R_0 = \frac{\beta}{\gamma}\label{eq:r0}
    \end{equation}

This quantity represents the number of individuals that are infected by a 
single individual in state I, when interacting with a totally Susceptible 
population. It is immediate to see that if $R_0$ is larger than 1 the 
infection will blossom. On the contrary, if this quantity is smaller than 1 
the infection dies out. \\

It should be kept in mind that once the process starts developing the $R_0$ 
should be replaced by the $R_e$ (\textit{i.e.} effective reproduction 
number).\\

\subsection{\label{sec:the_model_cyclic}The SEIR model for three groups and 
cyclical lock-down}

\begin{figure}
\centering
\includegraphics[scale=0.3]{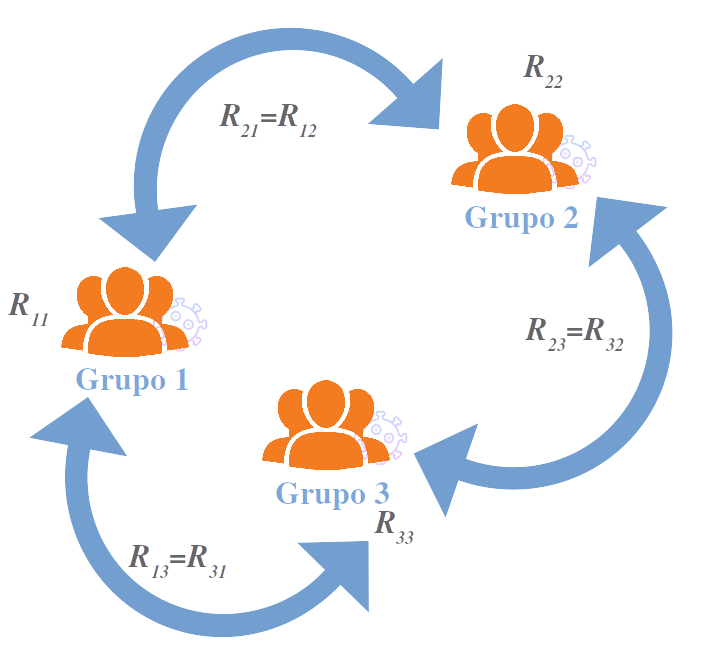}
\caption{\label{fig:esquema} Connectivity scheme for many groups. The 
disease spreading model corresponds to the SEIR one.  } 
\end{figure}

The SEIR model detailed in Section~\ref{sec:seir} assumes that the disease 
spreads over an homogeneous population. However, this may not be the case if 
some kind of grouping tendency exists among the individuals. The reproduction 
number may be different within the group, or, between groups. The latter will 
depend on the isolation degree among the groups. Two completely isolated 
groups attain independent evolutions. But a small infection leakage  povides 
the mean for the disease propagation from group to group.  
Fig.~\ref{fig:esquema} represents qualitatively this situation.\\

A multi-group SEIR model can mathematically represented as follows. Each group 
is labeled through indexes $i=1...N$. The former (scalar) states S,E,I,R 
become (column) arrays of states 

\begin{equation}
 (s_1,...,s_N)\, , (e_1,...,e_N)\, , (i_1,...,i_N)\, , (r_1,...,r_N)
\end{equation}

\noindent respectively. Notice that the total amount of individuals in each 
state corresponds to the sum of the elements of each array.\\

Since the propagation can actually occur within each group, or, among them, we 
define two sets of reproduction numbers. The first set corresponds to the 
reproduction number within group $R_{ii}=\beta_{ii}/\gamma_i$. The second set 
corresponds to the reproduction number between any two groups 
$R_{ij}=\beta_{ij}/\gamma_i$. The mixing among groups occurs as follows

\vspace{-4mm}

\begin{equation}
{\scriptsize
  \delta_{ijk}\,s_k\beta_{jl}i_l=
  \left(\begin{array}{ccc}
         s_1 & ... & 0 \\
         \vdots  & \ddots & \vdots \\
         0 & ... & s_N\\
        \end{array}\right)\,
  \left(\begin{array}{c}
         \beta_{11}i_1+...+\beta_{1N}i_N \\
          \vdots \\
         \beta_{N1}i_1+...+\beta_{NN}i_N\\
        \end{array}\right)
        }
\end{equation}

\vspace{2mm}

\noindent for $\delta_{ijk}$ representing the Kronecker tensor and 
$i,j,k,l=1,..., N$. The whole set of equations read
\vspace{-4mm}
 
\begin{equation}
 \left\{\begin{array}{lcl}
         \dot{s}_i(t) & = & -\delta_{ijk}\,\beta_{jl}\,s_k(t)\,i_l(t) \\
         & & \\
         \dot{e}_i(t) & = & \delta_{ijk}\,\beta_{jl}\,s_k(t)\,i_l(t) 
-\sigma_{ij}\,e_j(t)\\
         & & \\
         \dot{i}_i(t) & = & \sigma_{ij}\,e_j(t)-\gamma_{ij}\,i_j(t)\\
         & & \\
         \dot{r}_i(t) & = & \gamma_{ij}\,i_j(t)\\
        \end{array}\right.
\end{equation}

\vspace{-1mm}

\noindent for
$\sigma_{ij}=\mathrm{diag}(\sigma_{11},...,\sigma_{NN})$ and
$\gamma_{ij}=\mathrm{diag}(\gamma_{11},...,\gamma_{NN})$ (meaning that 
$\sigma_{ij}$ and $\gamma_{ij}$ are diagonal matrices).\\

\section{\label{simulaciones}Simulations}

We implemented the Runge Kutta 4th-order method in order to integrate the 
differential equations. The chosen time step was 0.1 (days).\\

As mentioned in Section~\ref{sec:modelo}, the parameters $\sigma$ and 
$\gamma$ represent the incubation rate and the recuperation rate, 
respectively. Therefore, $\sigma^{-1}$ and $\gamma^{-1}$ correspond to the 
mean incubation time and the mean recovering time, respectively. According to 
preliminary estimations for COVID-19, we consider the following parameter 
values for the SEIR model: $\sigma^{-1}=3\,$days and 
$\gamma^{-1}=4\,$days \cite{Milo,oms,hopkins,Lessler}.\\

\subsection*{\label{sec:ciclo} The cyclical work-lockdown implementation}

We first implemented a cyclic work-lockdown strategy in the same 
way as in Ref.~\cite{Alon}. This one corresponds to a cyclical schedule of 
lock-down and work days. As already mentioned in Section~\ref{sec:modelo}, ​the 
transmission parameter $\beta$ is the only parameter to be modified through 
lockdown policies. Recall that this parameter depends on the number of 
individual contacts. Therefore, policies like lock-down can reduce its 
value, and thus, the value of $R_0$ (see Eq.~\ref{eq:r0}).\\

The cyclical work-lockdown strategy corresponds to $W$-days of 
work and $L$-days of lock-down $(W, L)$. Two transmission 
parameters ($\beta$) may be accomplished on each period, due to the different 
number of contacts. These are $\beta_W$ for the working days and $\beta_L$
for the lock-down days, with the associated reproduction numbers $R_W$ and 
$R_L$, respectively (see Eq.~\ref{eq:r0}). We assume $R_L~=~$0.6 and 
$R_W~=~$1.5 (or greater) as in Ref.~\cite{Alon}.\\

\section{\label{sec:resultados}Results}

In this section we discuss the results obtained. We divided our investigation 
into two different scenarios. We first examined the case in which there is no 
intervention during the pandemic (Section~\ref{sec:sin_intervencion}), while 
the intervention case is left to Section~\ref{sec:con_intervencion}.\\

\subsection{\label{sec:sin_intervencion}The uncontrolled pandemic spreading}

\begin{figure}
\centering
\includegraphics[scale=1.0]{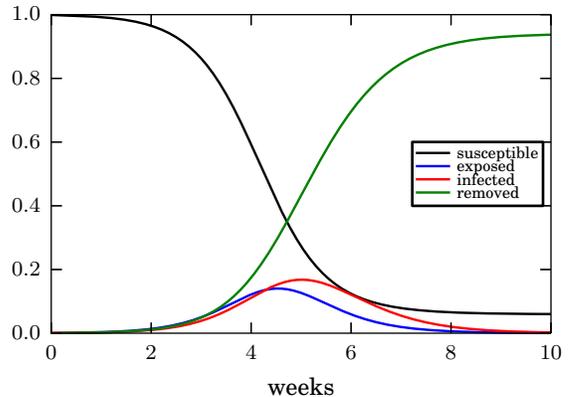}
\caption{\label{fig:seir_basico} Pandemic evolution. We used 
$\sigma^{-1}=3\,$days, $\gamma^{-1}=4\,$days and $\beta=$0.75 ($R_0$=3) 
\cite{Milo}. The initial condition was $s(0)=1, s(0)=0, i(0)=1\times 10^{-3}$ 
y $r(0)=0$.} 
\end{figure}

As a first step we examined how the pandemic spreads without 
any kind of intervention. In other words, withouth any cyclical strategy. 
According to Ref.~\cite{Milo}, we assumed that the reproduction number is 3. 
Fig.~\ref{fig:seir_basico} shows the evolution of the susceptible, exposed, 
infected and recovered individual along time. As can be seen, the number 
of susceptible individuals decreases monotonically during time. On the 
contrary, an opposite behavior can be observed for the recovered individuals. 
In this case, they increase monotonically along time.\\

Besides, Fig.~\ref{fig:seir_basico} shows that the exposed 
and infected individuals adopt a similar profile. As can be seen, the peak of 
infected individuals is reached at the fifth week. The height and width of the 
infected peak depends on the value of $R_0$. Thus, the greater the value of 
$R_0$, the higher the peak. And, in turn, narrower.\\

\subsection{\label{sec:con_intervencion}Evolution of the pandemic with 
cyclical strategy}

We now discussed the effects of applying the lockdown-work cycle during the 
pandemic. We will use the term homogeneous group to define 
a set of individuals that is in a single state. That is, a homogeneous group 
is one that is in a state of ``lockdown'' or ``work'', but not both. 
Two possible scenarios will be analyzed below:\\

\begin{itemize}
 \item An unique homogeneous group: there is only one 
homogeneous group.
\item Three homogeneous groups (multi-group): the crowd is divided into three 
groups of individuals of the same size. People in each group interact with 
those in 
the same group ($R_{ii}$). Also, people from one group can interact with  
people from another group ($R_{ij}$). It should be noted that, in this case, 
one group may be in a state of ``lockdown'', while the other two groups in a 
state of ``work'' or ``normal activity''.
\end{itemize}

\subsubsection{\label{sec:comparacion_israelies}The (4-8) and (4-10) cycles for 
a single group}

\begin{figure}[htbp!]
\centering
\subfigure[]{%
\includegraphics[scale=0.90]{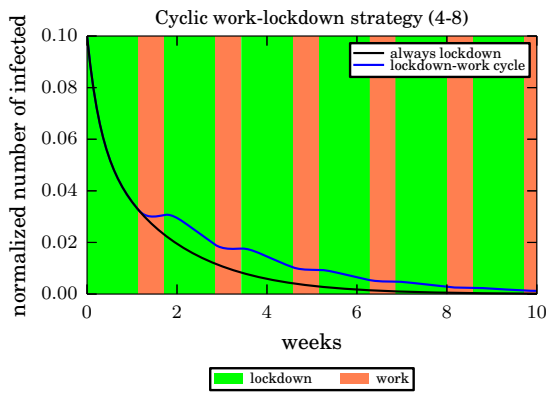}\label{
fig:4-8_varios}}
\subfigure[]{%
\includegraphics[scale=0.90]{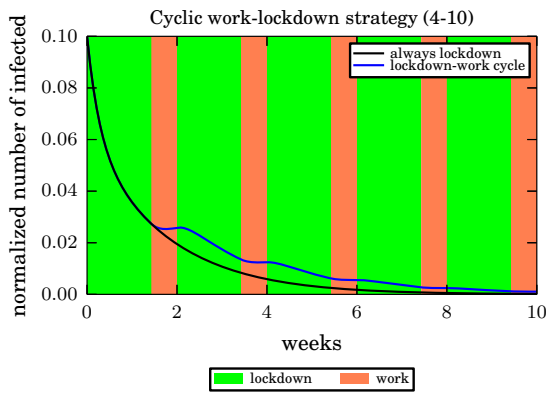}\label{
fig:4-10_varios}}
\caption[width=0.47\columnwidth]{\label{fig:infectados_vs_time_inicio} 
Normalized number of infected 
individuals along time. The initial condition was $s(0)=0.9, s(0)=0, s(0)=0.1$ 
and $s(0)=0$. The reproduction number for the ``always lockdown'' was 
$R_l=0.6$. On the other hand, in the ``lockdown-work cycle'' we used 
$R_w=1.5$ and $R_l=0.6$.} 
\end{figure}

Fig.~\ref{fig:infectados_vs_time_inicio} shows the number 
of infected individuals when we apply the lockdown-activity strategy for two 
different schedules (4-8 days and 4-10 days). Notice that this is applied after 
the number of infected surpasses $10\%$ of the total population. We compare 
this evolution with respect the case in which a permanent lockdown is 
applied (see Fig.~\ref{fig:infectados_vs_time_inicio}). As can be seen, in 
both cases, the best strategy in terms of the number of infected people, is a 
continuous (or strict) lock-down. In this case there is a monotonically 
decrease in the number of infected over time.\\

We can further see in Fig.~\ref{fig:infectados_vs_time_inicio} 
 that when applying the cyclical strategy (blue line) there is an 
increase in the number of cases due to the inclusion of the ``activity'' stage 
(equivalent to the release of the lock-down). In this sense, we can see that 
during this stage (see orange bars) the number of infected grows, unlike what 
happens in the ``lock-down'' stage (green bars).\\

Finally, we can observe a similar behavior when applying cycles 
(4-8) and (4-10) in terms of the number of infected. It should be noted that 
similar results are obtained in the case of the susceptible, exposed and 
recovered (not shown). Thus, we conclude that applying a cycle (4-8) 
is similar to the schedule (4-10). However, as we will see below, the first 
case allows us a continuous cycle of ``activity'' if three groups of 
individuals are considered. \\

In the next section we will focus on the study of the cycle (4-8).\\

\subsubsection{\label{sec:ciclo_grupo_simple}Analysis of the (4-8) estrategy 
for a single group}

\begin{figure}
\subfigure[]{\includegraphics[width=0.8\columnwidth]{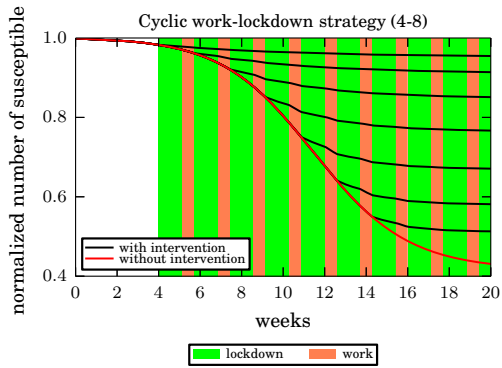}}
\subfigure[]{\includegraphics[width=0.8\columnwidth]{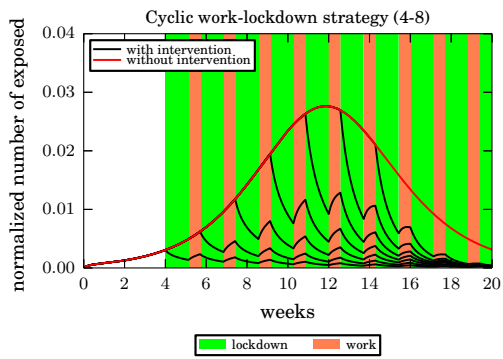}}
\subfigure[]{\includegraphics[width=0.8\columnwidth]{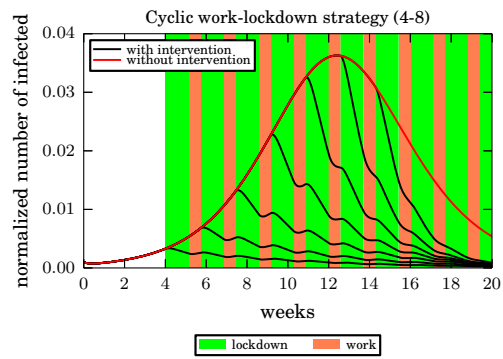}}
\subfigure[]{\includegraphics[width=0.8\columnwidth]{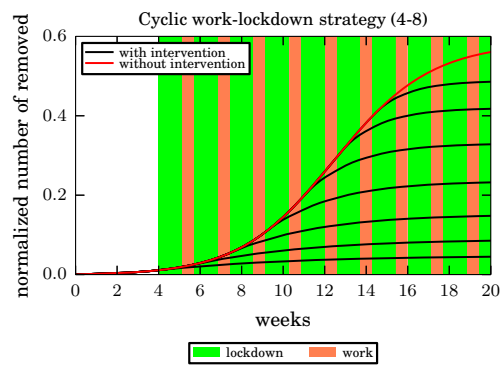}}
\caption{\label{fig:tipos_vs_time} Normalized number of susceptibles (a), 
exposed (b), infected (c) and recovered (d) as a function of time. The initial 
condition was $s(0)=1, e(0)=0, i(0)=1\times 10^{-3}$ y $r(0)=0$. The 
$R$ numbers for the intervention scenario were $R_w=1.5$ 
and $R_l=0.6$, while for ``non-intervention'' $R_0=$1.5.} 
\end{figure}

Unlike the previous case, we now analyze the effects of 
applying the lockdown-activity cycle during different stages of the pandemic. 
Fig.~\ref{fig:tipos_vs_time} shows the evolution of each stage 
(susceptible, exposed, infected and recovered) as a function of time. We also
plotted the case in which the system is allowed to evolve freely, 
equivalent to a ``continuous activity'' (red line). Starting from this curve, 
we applied the lockdown-activity cycle at different times of the evolution 
(see black lines). \\

As can be seen in Fig.~\ref{fig:tipos_vs_time}, the number of susceptible 
people decreases monotonically along time due to the progressive spreading of 
the virus. Also, we can note that once the lockdown-activity cycle has been 
implemented, the number of susceptible people reaches, seemengly  
in four weeks, an asymptotic regime. The same occurs with the recovered 
individuals (see Fig.~\ref{fig:tipos_vs_time}(d)), but in this case 
they increases monotonically with time.\\

Notice that the behavior of the number of susceptibles and 
recovered individuals is not affected by the start-up time. 
That is, the number of susceptibles decreases monotonically regardless of 
whether the population is in the lock-down or activity stage (the same occurs 
in the case of those recovered). However, we can observe a completely different 
behavior in the case of the exposed and the infected individuals. In this case, 
we can see that their behavior is affected according to the time of the 
evolution. Notice that the number of exposed individuals grows during the 
``activity'' stage. This can be explained if we take into account that during 
the ``activity'' stage, the frequency of contacts between susceptibles and 
infected increases (through $R_w$). Therefore, increases the probability of 
transmitting the virus to a susceptible. If this happens, the susceptible 
becomes to an exposed one. The opposite occurs in the lock-down stage, where 
there is a lower probability to transfer the virus.\\

Interesting, Fig.~\ref{fig:tipos_vs_time} also shows that the 
behavior of the number of infected individuals during the ``activity'' 
stage depends on the time of the intervention. That is, we can observe that, 
before week 11, the number of infected increases during the ``activity'' 
stage. Instead, this behavior is reversed (decreases) from, approximately, 
this point. So, the return to the activity stage does not affect the system in 
terms of an increase in the number of infected. Furthermore, it should be 
noted that this occurs before reaching the peak of the epidemic.\\ 

Up to now, we have analyzed how the epidemic evolves when 
considering a single homogeneous group, which can adopt a ``lock-down'' or 
``activity'' behavior. In the next Section we consider three groups of 
individuals. As previously mentioned, this case allows optimizing the strategy 
(4-8).\\

\subsubsection{\label{sec:ciclo_grupo_multiple} Analysis of the (4-8) strategy 
for three groups}

\begin{figure}
\centering
\includegraphics[width=\columnwidth]{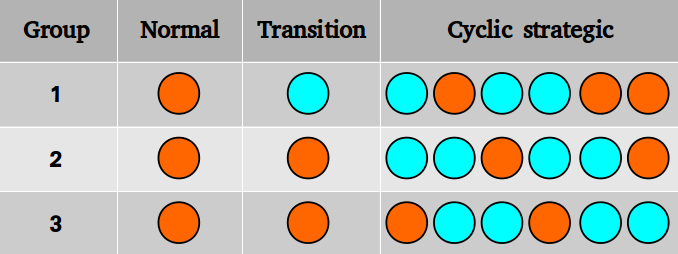}
\caption{\label{fig:cuadro} Schedule for the cycical estrategy on three 
groups. The orange and blue circles represent the working period and the 
lock-down period, respectively. The ``normal'' column means no estrategy at 
all. The column ``transition'' attains for the required shifting before the 
$4\times 8$ strategy is fulfilled.}
\end{figure}

We now focus on the population that appears splitted into three groups 
$i=1,2,3$. Recall that each group cycles through a normal working period and 
a lock-down period. The lock-down period fits into the mean time of the 
infection (say, $L=8$), while the working period lasts for approximately a 
working week ($W=4$). These $4\times 8$ periods for each group are set shifted 
in time, according to the schedule in Fig.~\ref{fig:cuadro}. \\

The reproduction number within each group is $R_{ii}=\beta_{ii}/\gamma_{ii}$. 
We assume two possible situations: normal working activity with
$R_{ii}=R_w>1$, or, the lock-down situation with $R_{ii}=R_l<1$. We further 
assume two possible reproduction values between the groups: a complete 
isolation ($R_{ij}=0$, $i\neq j$) or some leakage among them ($R_{ij}\ll 1$,  
$i\neq j$).\\

\begin{figure}
\centering
\subfigure[]{\includegraphics[scale=1.0]{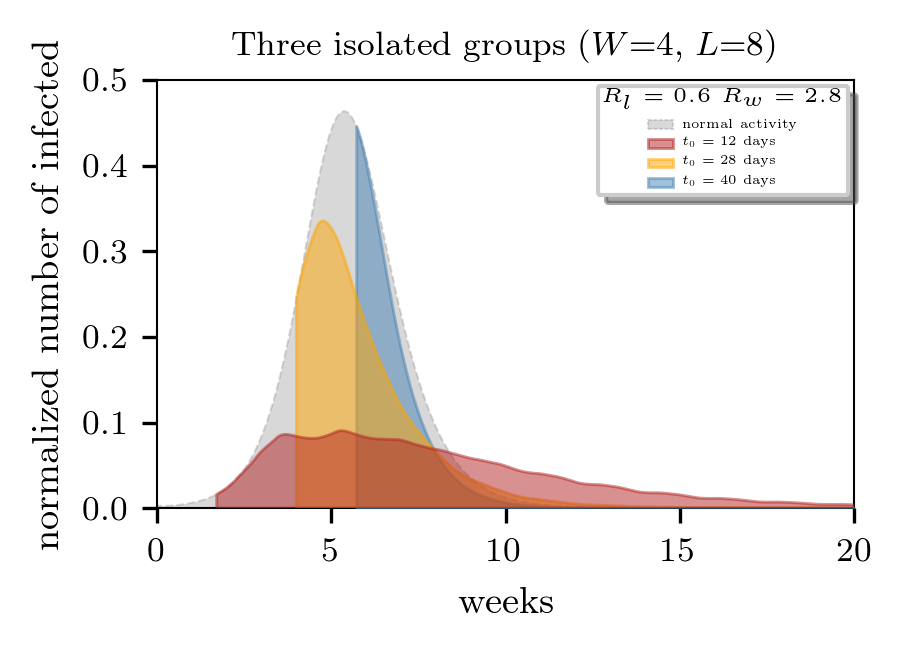}}
\subfigure[]{\includegraphics[scale=1.0]{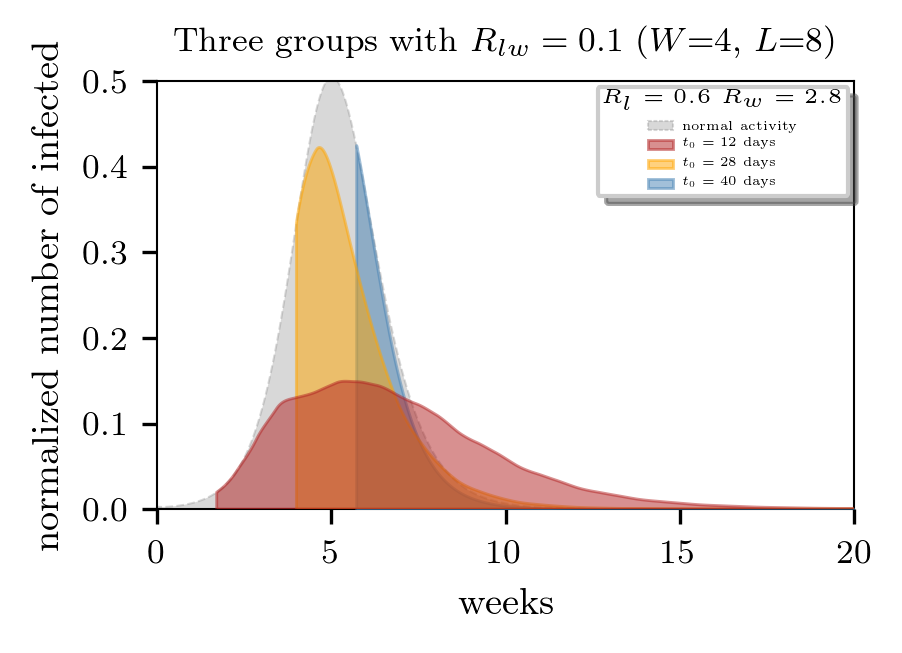}}
\caption{\label{fig:my_seir_groups_3_phase} 
Total number of infected people ($i_1+i_2+i_3$) vs. time. The initial 
conditions are $s(0)=1, e(0)=0, i(0)=1\times 10^{-3}$ and $r(0)=0$. The 
reproduction numbers for the ``normal working'' and the ``lock-down'' 
situations are $R_w=2.8$ and $R_l=0.6$, respectively. The working days are 
$W=4$, while the lock-down days are $L=8$. (a) The three groups are completely 
isolated ($R_{ij}=0$). (b) An infection leak exists between the group  
($R_{ij}=0.1$).} 
\end{figure}

Fig.~\ref{fig:my_seir_groups_3_phase} shows the time evolution for the total 
infected people for either isolated groups and non-insolated ones (see caption 
for details). The area in gray corresponds to the natural evolution with no 
strategy at all. The red, orange and blue colors correspond the $4\times 8$ 
strategy (see caption for details). The strategy start-up date is indicated in 
the legend. \\

\begin{figure}
\centering
\subfigure[]{\includegraphics[scale=1.0]{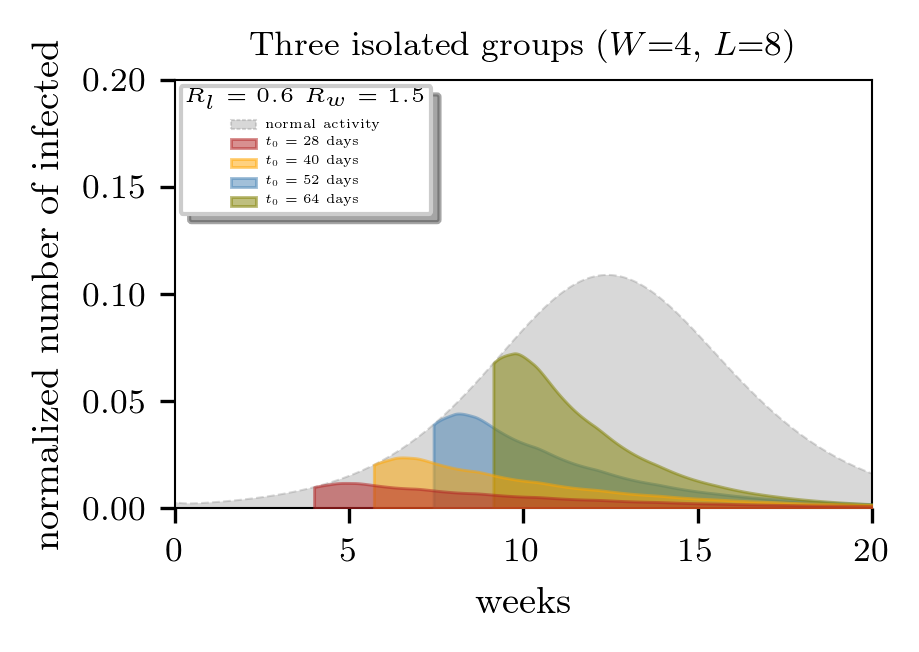}}
\subfigure[]{\includegraphics[scale=1.0]{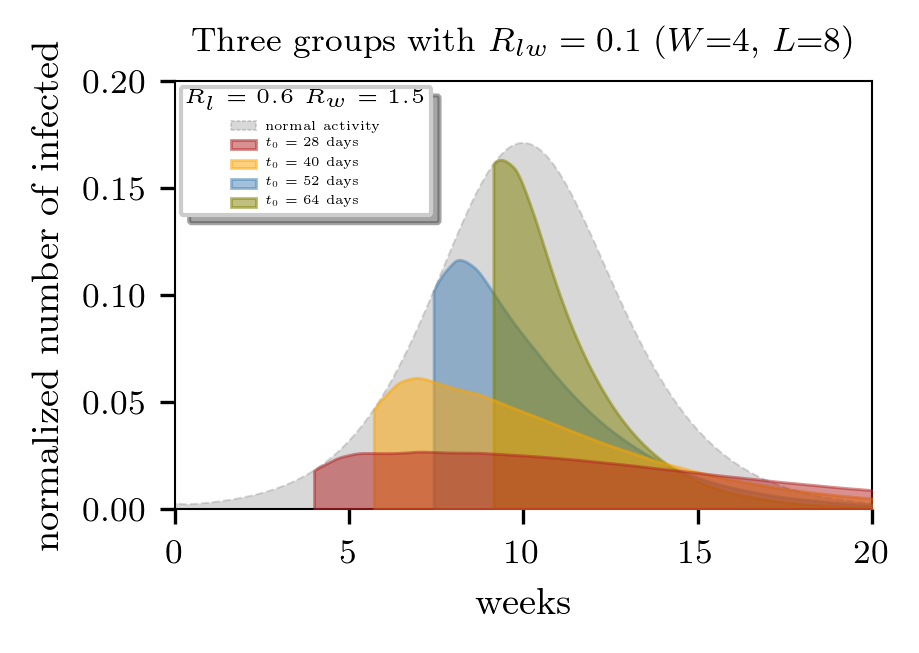}}
\caption{\label{fig:my_seir_groups_3_phase_R15} Total number of infected 
people ($i_1+i_2+i_3$) vs. time. The initial conditions are $s(0)=1, e(0)=0, 
i(0)=1\times 10^{-3}$ and $r(0)=0$. The reproduction numbers for the ``normal 
working'' and the ``lock-down'' situations are $R_w=1.5$ and $R_l=0.6$, 
respectively. The working days are $W=4$, while the lock-down days are $L=8$. 
(a) The three groups are completely isolated ($R_{ij}=0$). (b) An infection 
leak exists 
between the group ($R_{ij}=0.1$). } 
\end{figure}

The cyclical strategy decreases the amount of infected people as the process 
evolves, regardless of the infection leak among groups. However, if the groups 
remain completely isolated, the strategy exhibits a better performance. 
Besides, late implementation of the strategy provides poor results. \\  

Fig.~\ref{fig:my_seir_groups_3_phase_R15} shows similar results as in  
Fig.~\ref{fig:my_seir_groups_3_phase}, but for a weaker reproduction number. 
Notice that both scenarios decrease the total number of infected people. 
Thus, the $4\times 8$ strategy can be applied to a variety of infectious 
diseases (that is, different $R$).\\

\section{\label{conclusions}Conclusions}

We studied the time evolution of an infection describable by a SEIR 
compartmental model. Specifically, we implemented a cyclical asymmetric scheme 
by which the total population is divided in 3 parts, each performing a 4 
days normal activities period and an 8 days isolation period. This sequence is 
performed by each group comprising a third of the population in such a way 
that, at any given time, one third of the population is performing their usual 
``normal'' duties. In this way the solution of the SEIR equations indicate 
that the disease can be controlled while keeping a sensible degree of 
economical activity.\\

\begin{acknowledgments}
C.O.~Dorso is a Superior Researcher at National Scientific and Technological 
Council (spanish: Consejo Nacional de Investigaciones Cient\'\i ficas y 
T\'ecnicas - CONICET) and  Chief Professor at Depto. de F\'\i sica-FCEN-UBA. 
G.A.~Frank is an Assistant Researcher at CONICET. F.E.~Cornes es Lic. is a 
doctoral fellow at Depto. de F\'\i sica-FCEN-UBA. \\
\end{acknowledgments}

\end{document}